\documentstyle{EuroPhys}
\input EuroMacr
\input epsf.sty
\catcode`\@=11
\def\@evenhead{{\normalsize\bf \thepage}\hfil}
\def\rec#1#2{\gdef\@rec{}}
\def\@maketitle{\rm\vbox to0pt{}\vskip-34pt
 \parindent\z@
\def\EPLogo{\vtop{\hbox to \hsize
{{{\viiipt EUROPHYSICS LETTERS}\hfill {\normalsize\@data}}}}}
 \vglue 50pt
 {\Large\bf \@title\par\smallskip}
 \vskip 16pt \leftskip 25pt
 {\normalsize\@author }
 \vskip  4pt {\normalsize\it\@institute}
 \vskip 12pt {\normalsize\@rec}
 \vskip 16pt {\small\@pacs}
 \vskip 36pt
 \setcounter{page}{\value{startpage}}}
\catcode`\@=12
\begin{document}
\shorttitle{G.-L. INGOLD \etal THERMODYNAMICS OF NON-INTERACTING BOSONS ETC.}
\title{Thermodynamics of non-interacting Bosons in low-dimensional potentials}
\author{Gert-Ludwig Ingold\inst{1} 
\And Astrid Lambrecht\inst{2}\footnote{Present and 
permanent address: Laboratoire Kastler Brossel, Universit\'e Pierre et Marie 
Curie, Ecole Normale Sup\'erieure, Centre National de Recherche Scientifique, 
4~place Jussieu, BP74, F-75252 Paris Cedex 05, France}}
\institute{
     \inst{1} Institut f\"ur Physik, Universit\"at Augsburg, Memminger Str.~6, 
              D-86135 Augsburg,\\ Germany\\
     \inst{2} Max-Planck-Institut f\"ur Quantenoptik, Hans-Kopfermann-Str.~1,
              D-85748 Garching,\\ Germany}
\rec{}{}
\pacs{
\Pacs{03}{75Fi}{Phase coherent atomic ensemble (Bose condensation)}
\Pacs{05}{30Jp}{Boson systems}
\Pacs{64}{60-i}{General studies of phase transitions}}
\maketitle
\begin{abstract}
On the basis of a macroscopic ground state population it was argued
recently that Bose-Einstein condensation should occur in a one-dimensional
harmonic potential. We examine this situation by drawing analogies to Bosons 
in a two-dimensional box, where the thermodynamic limit is well-defined.
We show that in both systems although the ground state populations show sharp
onsets at the critical temperature, the behaviour of the specific heat is
analytic, which proves the absence of a phase transition in these systems.
\end{abstract}
The experimental study of ultracold trapped Bose gases \cite
{ander95,bradl95,davis95} has revived the interest in Bose-Einstein
condensation in the regime of weak or even vanishing interaction. In these
experiments the atoms are confined by external forces, which might be
modelled by a three-dimensional harmonic potential. Recently it was 
suggested on the basis of results for the
ground state population that in an effectively one-dimensional potential 
Bose-Einstein condensation
should occur \cite{kette96}. In the present paper we place the discussion
of non-interacting Bosons in a one-dimensional harmonic potential into a larger 
framework by drawing analogies to particles in a two-dimensional box with 
infinite walls, a system which has, in contrast to the  former one, the 
advantage that a well-defined thermodynamic limit exists. 
Within this framework we address the conceptual problem whether a macroscopic 
ground state population is a sufficient indicator for Bose-Einstein 
condensation to appear or if in contrast, thermodynamic quantities have to 
be considered.
In addition to the ground state population we study in particular the specific 
heat, calculated with a continuous density of states, which is permissible for 
large particle numbers. The result is confirmed
by an analysis based on the discrete level structure.

Bose-Einstein condensation \cite{londo38} is usually described within the
grandcanonical ensemble where the average number of particles is given by 
\begin{equation}
N=\int_0^\infty \drm E\rho (E)\frac z{\exp (\beta E)-z}.  \label{eq:n}
\end{equation}
Here, $z=\exp (\beta \mu )$ is the fugacity with $\beta =1/k_BT$ and the
chemical potential $\mu $. The energy scale is chosen such that the minimum
value of the $d$-dimensional external potential $U({\bf r})$ equals zero.
Then the classical density of states for particles of mass $m$ 
\begin{equation}
\rho (E)=\frac{1}{\Gamma (d/2)}\left(\frac{m}{2\pi \hbar^2}\right)^{d/2}\int 
\drm V_{{\rm cl}}\left[ E-U({\bf r})\right] ^{(d-2)/2},  \label{eq:rho}
\end{equation}
where the integration has to be taken over the classically accessible
region, vanishes for negative energies.
If $\rho(E)\to 0$ for small energies, $N$ reaches a finite maximum even for
the maximum value of the fugacity $z=1$. Additional particles are then found
in the ground state which is not included in the density of states (\ref
{eq:rho}). For a $d$-dimensional harmonic potential the density of states
varies like $E^{d-1}$ and therefore Bose-Einstein condensation is only
expected for $d=2$ and higher \cite{bagna87,bagna91}.

This approach has been criticized by various authors. For the
three-dimensional harmonic potential it was shown that corrections to the
density of states (\ref{eq:rho}) lead to corrections to the Bose
temperature, which are of order $N^{-1/3}$ \cite{gross95a}. Other work 
\cite{groot50,gross95,kirst96,brose96,haugs97} has addressed the 
question whether there are
effects due to the discreteness of the energy levels which is not included
in (\ref{eq:rho}). In this context, it was recently claimed that
Bose-Einstein condensation can be found even in the case of a
one-dimensional harmonic potential \cite{kette96}.

\begin{figure}
\begin{center}
\leavevmode
\epsfbox{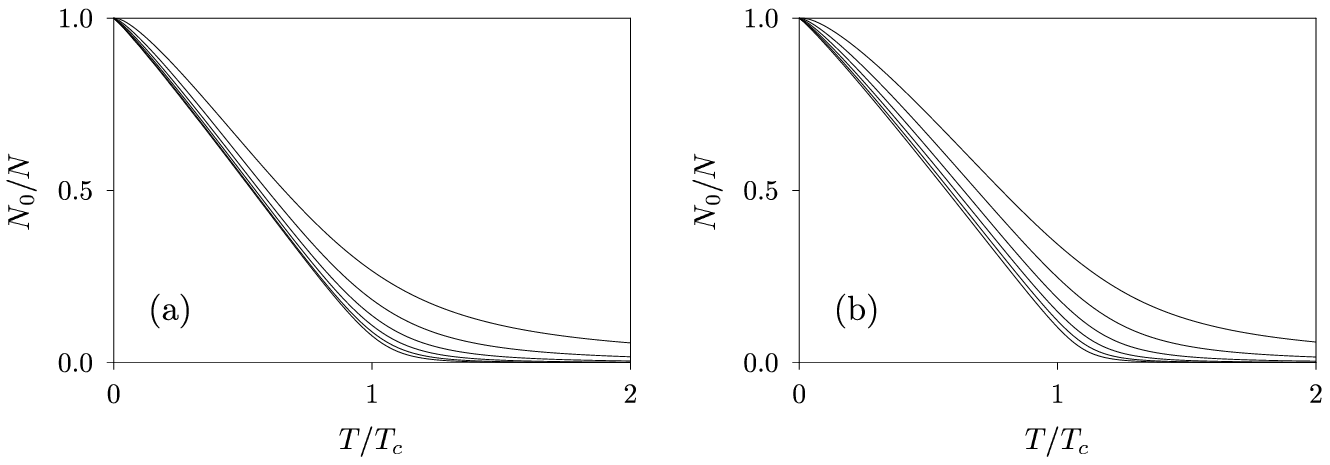}
\end{center}
\caption{Relative ground state occupation $N_0/N$ (a) for the one-dimensional 
harmonic potential and (b) for the two-dimensional box as a function of 
temperature $T$ scaled with the critical 
temperature $T_c$. The numerical results obtained by taking into account the 
discrete level structure correspond to $N=10^2, 10^3, 10^4, 10^5, 10^6,$ and 
$10^7$ particles from the upper right to the lower left curve.}
\label{fig:n01d}
\end{figure}

The argument given by Ketterle and van Druten is as follows. The relative
ground state occupation $N_0/N$ is calculated by taking into account the
discreteness of the energy levels. As an example, we show in 
fig.~\ref{fig:n01d}a results for particle numbers varying between $N=10^2$ and 
$N=10^7$ 
particles. The temperature is scaled with a
critical temperature determined by \cite{kette96} 
\begin{equation}
N=\frac{k_BT_c}{\hbar \omega }\ln \left( \frac{2k_BT_c}{\hbar \omega }\right) , 
\label{eq:tc}
\end{equation}
where $\hbar\omega$ is the energy level spacing. It can be seen that with
increasing number of particles $N$, the crossover between finite and
vanishing ground state occupation sharpens. This behaviour is similar 
to the three-dimensional
case, where Bose-Einstein condensation occurs, and it is thus concluded that
there is condensation even in one dimension.

However, in a one-dimensional harmonic potential with fixed frequency 
$\omega $ the critical temperature $T_c$ diverges in the limit $N\to \infty $. 
It is therefore not obvious how a phase transition in such a system may be
defined. Furthermore, a rescaling with $T_c$ is then equivalent to widening
the harmonic potential when increasing the particle number, i.e. lowering
the oscillation frequency towards zero (cf.\ eq.~(\ref{eq:tc})). 

We now address the question whether a macroscopic ground state population
is a sufficient indicator to define Bose-Einstein condensation in this
system. In this respect it is useful to consider another system with a
constant density of states $\rho (E)$. From eq. (\ref{eq:rho}) we find
for a two-dimensional box of length $L$
\begin{equation}
\rho (E)=\frac{mL^2}{2\pi \hbar ^2}.  \label{eq:rho2d}
\end{equation}
Here a proper thermodynamic limit $N,L\rightarrow \infty $, with
constant density $N/L^2$, can be performed and the notion of 
phase transition is well defined. While ($2\pi \hbar ^2/k_{\rm B}m)(N/L^2)$ 
provides the temperature scaling, either $N$ or $L^2$ determine how closely 
the thermodynamic limit is approached. For the two-dimensional box it is
generally accepted that, in contrast to the three-dimensional box, no
phase transition occurs (see \cite{groot50}, where the authors derive this 
conclusion, although the specific heat for the two-dimensional box is not 
correctly reproduced \cite{ziff77}).

An analogy to the one-dimensional harmonic oscillator may be drawn
by comparing its density of states $\rho (E)=1/\hbar \omega $ to
expression (\ref{eq:rho2d}). The temperature scale is now given by $N\hbar
\omega $ while the thermodynamic limit is reached either for $N\rightarrow
\infty $ or $\hbar \omega \rightarrow 0.$ For fixed frequency $\omega $ the
limit $N\rightarrow \infty $ leads to a change in the temperature scale. A
phase transition should therefore become apparent when rescaling the
temperature with $N$. For later purposes we note that the dimensionless
density for the two-dimensional box $N(\lambda _{\rm T}/L)^2,$ where 
$\lambda _{\rm T}=(2\pi \hbar ^2\beta /m)^{1/2}$ is the thermal de Broglie
wavelength, corresponds to $N\beta \hbar \omega $ for the one-dimensional
harmonic oscillator.

To check the analogy between the two systems, we have calculated the 
ground state population for the
two-dimensional box using discrete energy levels. Following the reasoning of
Ketterle {\it et al.} \cite{kette96} we define a critical temperature via
$N=2(L/\lambda _{\rm T,c})^2\ln(L/\lambda_{\rm T,c})$
and obtain the results shown in fig. \ref{fig:n01d}b. 
Clearly, as in the case of a one-dimensional harmonic potential, the 
ground state population for large $N$ shows a rather sharp transition
to macroscopic values below the critical temperature. No qualitative difference 
in the behaviour of the two systems is visible. The definition of 
Bose-Einstein condensation via a macroscopic ground state population 
enters therefore in conflict with predictions for the two-dimensional box. 
For a good criterion for condensation one instead has to resort to 
thermodynamic quantities.

To this purpose we consider in particular the specific heat, which recently 
has become accessible experimentally \cite{enshe96} 
\begin{equation}
C=\dif{}{T}\int_0^\infty \drm E\rho (E)\frac{Ez}{\exp
(\beta E)-z}.
\label{eq:c1}
\end{equation}
Here $C$ is expressed through the continuous density of states $\rho (E)$, 
which will prove to be sufficient for our purposes. Evaluating (\ref{eq:c1}) 
with constant density of states $\rho=1/\hbar\omega$ yields 
\begin{equation}
\frac C{Nk_{\rm B}}=\frac 2{N\beta \hbar \omega }\sum_{n=1}^\infty 
\frac{z^n}{n^2}+\frac 1{N\hbar \omega }\frac{\ln(1-z)}z
\dif{z}{\beta}
\label{eq:c2}
\end{equation}
and for the fugacity we obtain from (\ref{eq:n})
\begin{equation}
z=1-\exp (-\beta \hbar \omega N).  \label{eq:fuga}
\end{equation}
This expression is correct for high temperatures but breaks down when $z$
becomes of the order of its zero temperature value $N/(N+1)$, which is the
case at temperatures of the order of the critical temperature defined by
(\ref{eq:tc}). At these temperatures $z$ almost takes the value of one and
the difference to the approximation (\ref{eq:fuga}) is at most of order $1/N$.
Although (\ref{eq:fuga}) could not be used to determine the ground state
population at low temperatures, it allows us to obtain the correct result
for the specific heat. Corrections of order $1/N$ to the fugacity are
negligible in the first term on the right hand side of (\ref{eq:c2})
whereas they are potentially dangerous in the logarithmic term. However in
the temperature range of interest this term is strongly suppressed
because d$z/$d$\beta $ becomes exponentially small.

With (\ref{eq:fuga}) the specific heat takes the form
\begin{equation}
\frac C{Nk_B}=\frac 2y\sum_{n=1}^\infty \frac{(1-\exp (-y))^n}{n^2}-
\frac y{\exp (y)-1}  
\label{eq:cvf}
\end{equation}
with $y=\beta \hbar \omega N$. The particle number $N$ enters this result
only through a linear rescaling of the specific heat and the temperature.
A similar result has been derived for a canonical ensemble \cite{brose96}.
According to the above discussion the sum in (\ref{eq:cvf}) determines the
low temperature behaviour of the specific heat which is then found to be 
\begin{equation}
\frac C{Nk_B}=\frac{\pi ^2}3\frac 1y=\frac{\pi ^2k_{\rm B}}{3N}
\frac 1{\hbar \omega }T.  
\label{eq:cvlt}
\end{equation}
In view of the analogy between the one-dimensional harmonic oscillator and
the two-di\-men\-sion\-al box the specific heat of the latter one then becomes 
\begin{equation}
\left( \frac C{Nk_B}\right) _{\rm 2d box}=
\frac{\pi ^2k_{\rm B}}{3N}\frac{mL^2}{2\pi \hbar ^2}T.  
\label{eq:cvlt2d}
\end{equation}

\begin{figure}
\hbox to\textwidth{\vbox{\hbox to 0.5\textwidth{\epsfbox{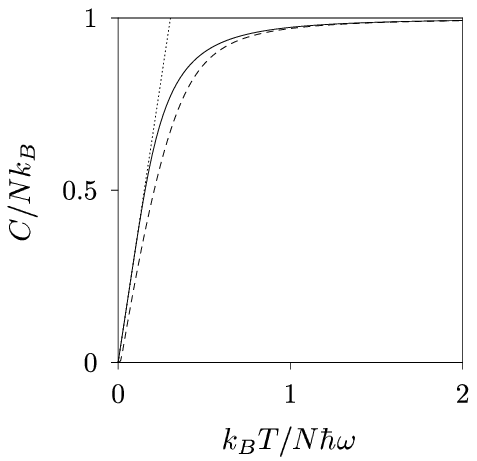}\hfill}}%
\vbox{\hbox to 0.5\textwidth{\epsfbox{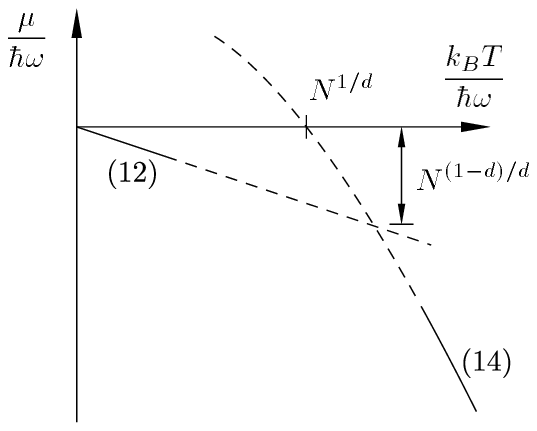}\hfill}}}
\hbox to\textwidth{\vbox{\hbox to 0.5\textwidth{\small Fig. 2.\hfill}}%
\vbox{\hbox to 0.5\textwidth{\small Fig. 3.\hfill}}}
\caption{Specific heat $C$ of a Bose gas in a one-dimensional harmonic
potential as a function of temperature $T$. The full line represents the
result of a calculation with constant density of states with 
its low temperature approximation (dotted line), while the dashed
line is for $N=10$ particles taking into account the discrete level
structure. For $N=10$ eq.~(\protect\ref{eq:tc}) yields 
$k_BT_c/N\hbar \omega =0.454$. Similar numerical results have been
obtained in \protect{\cite{brose96,haugs97}}.}
\label{fig:cv}
\caption{Illustration of the reasoning leading to (\protect\ref{eq:cptb}). 
The lines correspond to the upper bound of the chemical potential 
(\protect\ref{eq:cplow}) and
the high temperature approximation (\protect\ref{eq:cphigh}), respectively.}
\label{fig:approx}
\end{figure}

In fig.~\ref{fig:cv}, expression (\ref{eq:cvf}) is depicted by the full line
with its low temperature approximation (\ref{eq:cvlt}). The dashed line shows the result of a numerical calculation for 10
particles in a one-dimensional harmonic potential taking into account the
discreteness of the spectrum. Even 
for such small $N$, (\ref{eq:cvf}) gives a rather good approximation. For
100 or more particles the difference between the result for the
one-dimensional harmonic oscillator based on the discrete level structure on
one hand and (\ref{eq:cvf}) on the other hand becomes negligible. For the
two-dimensional box the convergence of the calculation using discrete
energies to the result for a constant density of states is somewhat slower
due to the higher dimensionality of the system.

Thus for large particle numbers the form of the specific heat becomes
independent of $N$ and neither singularities build up in the thermodynamic
limit $N\to \infty $ nor a maximum appears at the critical temperature like the one found in the result sketched in \cite{groot50}. Furthermore, the limiting result 
(\ref{eq:cvf}) is analytic for finite temperatures and we conclude that, as in 
the two-dimensional box,
there is no phase transition in a one-dimensional harmonic potential.

The difference between the behaviour of a non-interacting Bose gas in a
one-dimensional and higher-dimensional harmonic potentials may also be
understood in terms of discrete energy levels. An appropriate criterion for
the possibility of Bose-Einstein condensation to occur is whether in the
thermodynamic limit $N\to\infty$ the chemical potential equals the ground
state energy below the critical temperature or not. In the following we will
show that this criterion is not met in a one-dimensional harmonic potential.
We start out with the average number of particles in a grand canonical
description 
\begin{equation}
N=\frac{1}{\exp(-\beta\mu)-1}+\sum_{n=1}^{\infty}
\frac{g_n^{(d)}} {\exp[\beta(\hbar\omega n-\mu)]-1},  
\label{eq:ndisc}
\end{equation}
where $g_n^{(d)}$ is the degeneracy of the $n$-th eigenlevel of the $d$%
-dimensional harmonic potential. The first term on the right-hand-side of 
(\ref{eq:ndisc}) represents the population of the ground state.

One derives an upper bound of the chemical potential at arbitrary
temperatures by considering a situation where the excited states do
not contribute to the number of particles 
\begin{equation}
\frac{\mu}{\hbar\omega}\le-\frac{1}{N}\frac{k_BT}{\hbar\omega}.
\label{eq:cplow}
\end{equation}
Naturally this case coincides with the low temperature limit $%
\beta\hbar\omega\gg 1$. Chemical potential and temperature are taken with
respect to the energy level spacing $\hbar\omega$ of the harmonic potential.
This is the appropriate choice if $\omega$ is kept fixed in the
thermodynamic limit.

For high temperatures the ground state population becomes very small and
therefore $\beta\vert\mu\vert\gg 1$. We may then approximate expression 
(\ref{eq:ndisc}) for the average particle number by 
\begin{equation}
N=\exp{(\beta\mu)}\sum_{n=0}^{\infty}g_n^{(d)}\exp(-\beta\hbar\omega n)=
\frac{\exp(\beta\mu)}{(1-\exp(-\beta\hbar\omega))^d}.
\label{eq:napprox1}
\end{equation}
Except for very low particle numbers the high temperature regime implies 
$\beta\hbar\omega\ll 1$. The chemical potential is then found as 
\begin{equation}
\frac{\mu}{\hbar\omega}=
\frac{\ln\left[N(\beta\hbar\omega)^d\right]}{\beta\hbar \omega}  
\label{eq:cphigh}
\end{equation}
and equals zero at $k_BT/\hbar\omega=N^{1/d}$.
Since the ground state energy was taken to be zero, the high temperature
result suggests that there might be Bose-Einstein condensation below a
temperature $\hbar\omega N^{1/d}/k_B$. A more detailed analysis leads
to a prefactor of the order of one.

To exclude the existence of a phase transition in a one-dimensional
harmonic potential, it is sufficient to consider the upper bound (\ref
{eq:cplow}) for the chemical potential at its intersection with the high
temperature expression (\ref{eq:cphigh}). There the value of the chemical
potential becomes (cf.~fig.~\ref{fig:approx}) 
\begin{equation}
\frac{\mu}{\hbar\omega}=-N^{(1-d)/d}.  \label{eq:cptb}
\end{equation}
For $d=1$ this is of order one independently of the particle number $N$ and
one expects a smooth crossover at the intersection point. This behaviour
forbids a phase transition in the one-dimensional case.
On the other hand, for $d\ge 2$, the chemical potential at the intersection
vanishes for $N\to\infty$ opening the possibility of a sharp transition at a
temperature of order $N^{1/d}$, below which the chemical potential strictly
equals the ground state energy. We thus recover the predictions of the
theory based on the density of states (\ref{eq:rho}). 

In conclusion, by studying the specific heat and using an analogy with
two-dimensional box systems we have shown that a macroscopic 
ground state population alone is not a sufficient indicator of a phase 
transition. In higher dimensional systems which show a condensation 
phenomenon a macroscopic ground state population is always accompanied by some 
signature in the specific heat or higher derivatives of the free energy. Their 
absence for the case of a one-dimensional harmonic potential leads us to 
conclude that no phase transition occurs in this system. The main interest of 
Bose-Einstein condensation lies in the particular coherence properties of 
condensates. Then the question arises to which extent the ground state
population can give information about coherence. The results of this paper show 
that it might be necessary, maybe not sufficient, to carefully study 
thermodynamic quantities as well.

\stars

\end{document}